\documentclass[10pt,twocolumn,letterpaper]{article}

\usepackage[pagenumbers]{cvpr} % To force page numbers, e.g. for an arXiv version

\usepackage{graphicx}
\usepackage{amsmath}
\usepackage{amssymb}
\usepackage{booktabs}

\usepackage[pagebackref,breaklinks,colorlinks]{hyperref}

\usepackage[capitalize]{cleveref}
\crefname{section}{Sec.}{Secs.}
\Crefname{section}{Section}{Sections}
\Crefname{table}{Table}{Tables}
\crefname{table}{Tab.}{Tabs.}

\def\cvprPaperID{*****} % *** Enter the CVPR Paper ID here
\def\confName{CVPR}
\def\confYear{2023}

\begin{document}

\title{Recyclable Semi-supervised Method Based on Multi-model Ensemble for Video Scene Parsing}
\author{
Biao Wu\and
Shaoli Liu\and
Diankai Zhang\and
Chengjian Zheng\and
Si Gao\and
Xiaofeng Zhang\and
Ning Wang\and
State Key Laboratory of Mobile Network and Mobile Multimedia Technology,ZTE,China
\and
{\tt\small \{wu.biao,liu.shaoli,zhang.diankai,zheng.chengjian,gao.si,zhang.xiaofeng18,wangning\}@zte.com.cn}
}
\maketitle

%%%%%%%%% ABSTRACT
\begin{abstract}
   Pixel-level Scene Understanding is one of the fundamental
problems in computer vision, which aims at recognizing
object classes, masks and semantics of each pixel in the
given image. Since the real-world is actually video-based
rather than a static state, learning to perform video
semantic segmentation is more reasonable and practical for
realistic applications. In this paper, we adopt
Mask2Former as architecture and ViT-Adapter as
backbone. Then, we propose a  recyclable semi-supervised
training method based on multi-model ensemble. Our method
achieves the mIoU scores of 62.97\% and 65.83\% on
Development test and final test respectively. Finally, we
obtain the 2nd place in the Video Scene Parsing in the Wild
Challenge at CVPR 2023.
\end{abstract}

%%%%%%%%% BODY TEXT
\section{Introduction}
\label{sec:intro}

Semantic segmentation is a dense pixel prediction task that assigns a semantic label to each pixel in an image or video, such as person, car, building, etc. Thanks to the rapid development of deep learning and the availability of various semantic segmentation datasets, significant progress has been made in image semantic segmentation. However, the real world is dynamic and constantly changing, and it is more practical to learn how to perform video semantic segmentation for realistic applications rather than relying solely on static images. Video semantic segmentation is even more challenging. It requires considering the relationship between the temporal and spatial dimensions, as well as tackling issues such as object motion, occlusion, and background variations in videos.

Many studies have focused on exploiting the inter-frame redundancy of videos, such as using optical flow information, spatio-temporal attention networks, etc., which can better consider the temporal and spatial relationships in videos and improve the performance of video semantic segmentation. In addition to model structure advancements, the quality and scale of datasets have also received increasing attention. The Video Scene Parsing in the Wild(VSPW) dataset \cite{miao2021vspw} is the first attempt to tackle the challenging video scene parsing task by considering diverse scenarios, which covers a wide range of real-world scenarios (e.g., art galleries, lecture rooms, beach, and street views) and categories from both things (e.g., person, car, desk) and stuff (e.g., road, wall, sky). Furthermore, joint learning and weakly supervised learning have become hot topics in recent years. Joint learning can combine image and video data to improve the performance of video semantic segmentation, while weakly supervised learning can train models with fewer labeled data, reducing annotation costs.

In this paper, we propose the recyclable semi-supervised method\cite{chen2021semi}\cite{hung2018adversarial} based on multi-model ensemble. We finally obtain the scores of 62.97\% and 65.83\% mIoU on test\_part1 and testing set of VSPW, respectively. And our solution is ranked second place on PVUW2023 Video semantic segmentation challenge at CVPR2023.

\begin{figure*}[t]
  \centering
  %\fbox{\rule{0pt}{2in} %
   \includegraphics[width=1.0\linewidth]{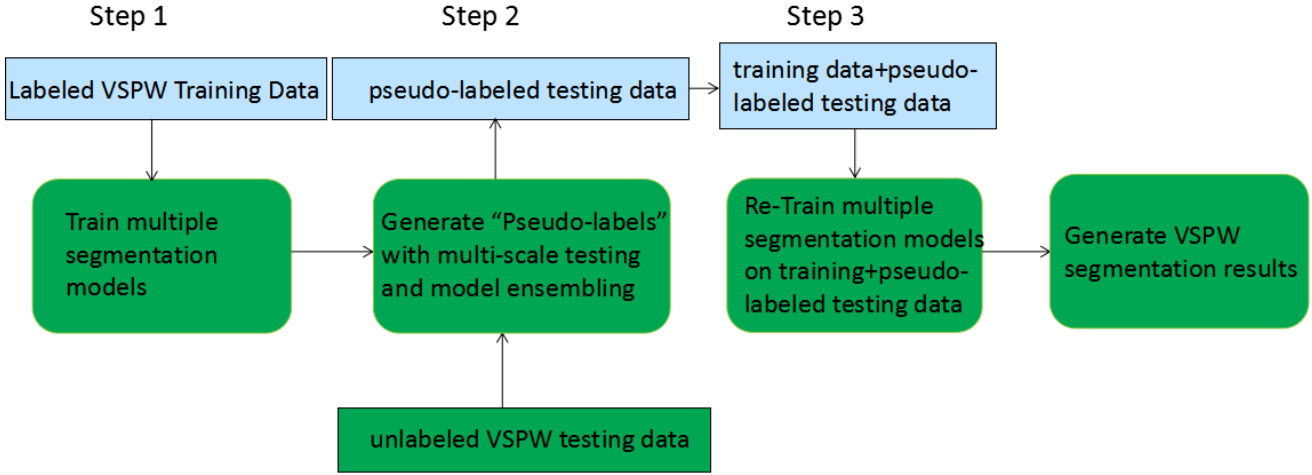}  %}
   \caption{The overview pipeline of our method.}
   \label{fig:onecol}
\end{figure*}

%-------------------------------------------------------------------------
\section{Method}
\label{sec:method}
In this section, we first describe the overview pipline of our network. And then, we introduce our Recyclable Semi-supervised method based on multi-model ensemble. Finally, we describe the ensemble strategy.

\subsection{Overview}

We use the masked attention mask transformer (Mask2Former)\cite{cheng2022masked} as the  image segmentation architecture,  and chose ViT-Adapter\cite{chen2022vision} as the  backbone.

\subsection{Backbone Structure}

Masked image modeling (MIM) has demonstrated impressive results in self-supervised representation learning by recovering corrupted image patches. BEiTv2\cite{peng2022beit} is a method based on MIM, so we leverage BEiTv2 in ViT-Adapter, in which we set embed\_dim=1024, depth= 24, mlp\_ratio = 4, num\_heads= 16.

\subsection{Recyclable Semi-supervised method based on multi-model ensemble}
As illustrated in Figure 1, in this competition how to using the unlabeled test data is crucial, so we using semi-supervised approach. Specificly, in the first step we use the labeled training data to train two models with different parameter configuration, and then we use different image ratios  for multi-scale testing, and apply horizontal flipping for each scale. Moreover, we ensemble the two different models based on the class probability for boosting the performance, we directly using this result as the pseudo label of VSPW test data.  In the third step, we use the VSPW training and the pseudo-labeled testing data to retrain the model. These steps can be repeated to obtain better testing  results.

%-------------------------------------------------------------------------

\subsection{Pseudo-labeling strategy}
Inspired by semi-supervised learning, we adopt an pseudo-labeling strategy to make full use of  the VSPW testing dataset.

In particular, we first build two network to provide a target probability for each of N classes for every pixel in
each image. The two networks have the same network structure but with different parameter configurations (training setting, crop size, etc) to ensure the effective of model aggregation. And then we adopt an emsemble strategy based on probability. In detail, for a given pixel, we select the top class prediction of the two network as the pseudo label. In order to make pseudo label have higher
confidence, we threshold the label. Those pixels whose maximum prediction probability exceed the threshold will be set as true label, otherwise the pixels will be marked as “ignore” class. In our experiments, the threshold is set to 0.4.

\subsection{Loss}
We jointly use the Dice Loss\cite{li2019dice} and CrossEntropy Loss to improve the global and local segmentation performance.

%------------------------------------------------------------------------
\section{Experiments}
\label{sec:Experiments}

In this part, we will describe the implementation details of our proposed method and report the results on the PVUW2023  challenge final test set.

\begin{table}
  \centering
  \resizebox{1.0\columnwidth}{!}{
  \begin{tabular}{c|cccc}
    \toprule
    Method & Backbone &mIoU &  VC8 & VC16 \\
    \hline
Mask2Former&Swin-L	&0.5709	&0.8876	&0.8596\\
Mask2Former&BEiT-L	&0.5854	&0.8964	&0.8611\\
Mask2Former&ViT-Adapter-L&0.6140&0.9007	&0.8638\\
    \bottomrule
  \end{tabular}
  }
  \caption{Experiments of different backbones on PVUW2023 challenge test part 1.}
  \label{tab:1}
\end{table}

\begin{table}
  \centering
  \resizebox{1.0\columnwidth}{!}{
  \begin{tabular}{c|cccc}
    \toprule
    Method&Crop Size & mIoU & VC8 & VC16 \\
    \hline
ViT-Adapter-L&224	&0.5795	&0.9044	&0.8695  \\
ViT-Adapter-L&320	&0.5890	&0.9051	&0.8715  \\
ViT-Adapter-L&640	&0.6140	&0.9007	&0.8638  \\
ViT-Adapter-L&768	&0.6169	&0.9056	&0.8723  \\
    \bottomrule
  \end{tabular}
  }
  \caption{Experiments of different crop size with ViT-Adapter-L on PVUW2023 challenge test part 1.}
  \label{tab:2}
\end{table}

\begin{table*}
  \centering
  \begin{tabular}{cc|ccc|ccc}
    \toprule
    Method & crop size & multiscale & flip & Semi-supervised training &	mIoU	&VC8	&VC16 \\
    \hline
    ViT-Adapter-L    &640&\checkmark &\checkmark & 	&0.6207	&0.9343	&0.9112 \\
    $\star$ ViT-Adapter-L &640&\checkmark &\checkmark &\checkmark &0.6238	&0.9330	&0.9096 \\
    ViT-Adapter-L    &768&\checkmark &\checkmark & 	&0.6236	&0.9331	&0.9097 \\
    $\star$ ViT-Adapter-L &768&\checkmark &\checkmark &\checkmark &0.6276	&0.9432	&0.9214 \\
    Ensemble($\star$)&   &           &           &           &0.6297	&0.9316	&0.9051 \\
    \bottomrule
  \end{tabular}
  \caption{Experiments of Inference Augmentation on PVUW2023 challenge test part 1.}
  \label{tab:example}
\end{table*}

%-------------------------------------------------------------------------
\subsection{Datasets}

The VSPW Dataset annotates 124 categories of real-world scenarios, which contains 3,536 videos, with 251,633 frames totally. Among these videos, there are 2806 videos in the training set, 343 videos in the validation set, and 387 videos in the testing set. In order to enrich our training samples, both the training and validation set are used for training. The model based on Transformers has a large number of parameters, and  increasing the number of training samples is beneficial for performance improvement. We introduce additional data to train our model, such as the ADE20k\cite{zhou2017scene} and COCO\cite{lin2014microsoft} datasets. The COCO dataset is used in the pre-training phase. The COCO and ADE20k dataset labels are mapped to the VSPW dataset through label remapping, and categories that do not exist in the VSPW dataset are marked as 255.

\subsection{Training Configuration}
All models in our experiments are implemented in the Pytorch\cite{paszke2017automatic} framework. For data augmentation, we perform random resizing within ratio range [0.5, 2.0], random cropping, random horizontal flipping, and color jitter on images. During the training phase, the backbone of our model is pretrained on the ImageNet22K\cite{deng2009imagenet} dataset. An AdamW optimizer is applied with the initial learning rate of 1e-5, beta = (0.9, 0.999), and the weight decay of 0.05. A linear warmup is used. To prevent overfitting, the training iterations is set to 80k.

\begin{table}
  \centering
  \resizebox{1.0\columnwidth}{!}{
  \begin{tabular}{ccccc}
    \toprule
    Team & mIoU & Weight IoU & VC8 & VC16 \\
    \midrule
    minbao&	0.6595& 0.7514& 0.9108& 0.8783  \\
    SiegeLion& 0.6583& 0.7512& 0.9432& 0.9214 \\
    csj& 0.6484& 0.7395& 0.9198& 0.8903 \\
    naicha& 0.6475& 0.7388& 0.9168& 0.8868 \\
    chenguanlin& 0.6372& 0.7351& 0.8984& 0.8619 \\
    SUtech& 0.5889& 0.7097& 0.9336& 0.9041 \\
    \bottomrule
  \end{tabular}
  }
  \caption{Comparisons with other teams on the PVUW2023 challenge final test set.}
  \label{tab:example}
\end{table}

\subsection{Ablation Studies}
Recently, many studies have shown that models based on transformer exhibit strong feature extraction capabilities in tasks such as dense detection and segmentation. Therefore, we explore the application of models based on transformer in video semantic segmentation tasks. The experimental results of different backbones and methods are shown in Table 1. From the table, it can be seen that selecting ViT-Adapter-L as the backbone has significantly better performance than Swin-L\cite{liu2021swin} and BEiT-L\cite{bao2021beit}. In the subsequent experiments, we selected ViT-Adapter-L as the backbone network. The input resolution of the model is also a key factor affecting model performance, so we explore the crop size of the network input. From the table, we can find that the larger the input resolution of the model, the better the performance tendency of the model. Due to computational resource constraints, we choose the input resolution of the model of  640*640 and 768*768.

\subsection{Semi-supervised training}
Semi-supervised learning aims to improve the performance on unlabeled dataset.
Inspired by this, we use the pseudo-labeling strategy to further expand the training samples.
We first perform multi-scale and horizontal flipping operations on two models with crop sizes of 640 and 768, and then ensemble the results of these two models as the results of the teacher network.
We use the results of the above teacher network as pseudo labels.
We combine pseudo label data with the original training dataset to form a new dataset.
We continue to refine the model on the new dataset for multiple times.
For pseudo label, we use hard thresholds to filter the pixel values of pseudo labels for the results of the teacher network.
If the probability value corresponding to the pixel value is higher than 0.4,
it is considered to be a reliable pixel value,
which is involved in the calculation of the loss function, and other pixel values are ignored.

\subsection{Inference Augmentation}

Inference augmentation is an effective method for improving the performance of models. We achieve better segmentation performance by using multi-scale and horizontal flipping for each scale where the selected scales are [512./896., 640./896., 768./896., 896./896., 1024./896., 1152./896., 1280./896., 1408./896.]. From Tables 2 and 3, it can be seen that The multi-scale and horizontal flipping results increase the mIoU indicator by 0.6 percentage points compared to the single scale results. In order to further improve model performance, we save the soft classification results (a matrix with shape 1 * h * w) of multi-scale and horizontal flipping, and then ensemble the soft classification results of two models with cropping sizes of 640 and 768. From Table 3, Semi-supervised training and model ensemble improved the mIoU by approximately 0.4 percentage points and 0.2 percentage points respectively. Finally, we obtain the 2nd place in the final test set, as shown in Table 4.

\section{Conclusion}
\label{sec:Conclusion}

In this paper, we propose a Recyclable Semi-supervised Method Based on Multi-model Ensemble for Video Scene Parsing. We Propose a ensemble method to get more accurate probability by fusing the results of different models, and employ recyclable semi-supervised method to make full use of unlabeled and labeled data for enriching the training datasets.

With the proposed method, Our solution not only performs well in mIoU, but also outperforms others significantly in video frame continuity and consistency on the PVUW2023 challenge.

%-------------------------------------------------------------------------
\section*{Acknowledgements}
\label{sec:Acknowledgements}
Thanks for colleagues from other departments for providing the cloud server for training.

%%%%%%%%% REFERENCES
{\small
\bibliographystyle{unsrt}
\bibliography{Recyclable_Semi_supervised_Method_Based_on_Multi_model_Ensemble_for_Video_Scene_Parsing}
}

\end{document}